\documentclass[journal]{IEEEtran}

\usepackage{fancyhdr}
\usepackage{makeidx}
\usepackage{graphicx}      % include this line if your document contains figures
\usepackage{multicol}
\usepackage{amsmath}
\usepackage{cite}
\usepackage{booktabs}
\usepackage{epstopdf}
\usepackage{url}
\usepackage{bm}
\usepackage{afterpage}
\usepackage{subfigure}
\usepackage{multirow}
\usepackage{array}
\usepackage[strings]{underscore}
\makeatletter
\newtheorem{theorem}{Theorem}
\newtheorem{prepos}{Proposition}

\DeclareMathAlphabet{\mathpzc}{OT1}{pzc}{m}{it}
\def\qqquad{\hskip3em\relax}
\newcommand*{\rom}[1]{\expandafter\@slowromancap\romannumeral #1@}
\makeatother
\begin{document}

\title{A Learning-based Stochastic MPC Design for Cooperative Adaptive Cruise Control to Handle Interfering Vehicles}
\author{Hadi Kazemi, Hossein Nourkhiz Mahjoub, Amin Tahmasbi-Sarvestani, Yaser P. Fallah
	\thanks{This work is supported by the National Science Foundation under CAREER
	Grant Number 1664968.}
	\thanks{Hadi Kazemi and Amin Tahmasbi-Sarvestani are PhD students with the
	Department of Electrical Engineering and Computer Science, West Virginia
	University, Morgantown, WV 26506 USA (e-mails: hakazemi@mix.wvu.edu
	, amtahmasbi@mix.wvu.edu).}
	\thanks{Hossein Nourkhiz Mahjoub is a PhD student with the Department of Electrical
	and Computer Engineering, University of Central Florida, Orlando, FL 32826
	USA (e-mail: hnmahjoub@knights.ucf.edu).}
	\thanks{Yaser P. Fallah is an associate professor with the Department of Electrical and
	Computer Engineering, University of Central Florida, Orlando, FL 32826 USA
	(e-mail: yaser.fallah@ucf.edu).}}

\maketitle
\thispagestyle{fancy}
\begin{abstract}
Vehicle to Vehicle (V2V) communication has a great potential to improve reaction accuracy of different driver assistance systems in critical driving situations. Cooperative Adaptive Cruise Control (CACC), which is an automated application, provides drivers with extra benefits such as traffic throughput maximization and collision avoidance. CACC systems must be designed in a way that are sufficiently robust against all special maneuvers such as cutting-into the CACC platoons by interfering vehicles or hard braking by leading cars. To address this problem, a Neural- Network (NN)-based cut-in detection and trajectory prediction scheme is proposed in the first part of this paper. Next, a probabilistic framework is developed in which the cut-in probability is calculated based on the output of the mentioned cut-in prediction block. Finally, a specific Stochastic Model Predictive Controller (SMPC) is designed which incorporates this cut-in probability to enhance its reaction against the detected dangerous cut-in maneuver. The overall system is implemented and its performance is evaluated using realistic driving scenarios from Safety Pilot Model Deployment (SPMD). 
\end{abstract}

\IEEEpeerreviewmaketitle
%===============================================================================

\section{Introduction}
\IEEEPARstart{D}{rivers} are the most important and influential entities of non-autonomous vehicles in ground Intelligent Transportation System (ITS). 
A revolutionary age of modern driving has been initiated by the advent of safety and
comfort driving applications that aim at assisting drivers in vehicle control. Forward collision warning \cite{1,2,3,4}, lane keep assistance \cite{5,6,7}, automatic braking \cite{8}, adaptive cruise control \cite{9, 10}, efficiency \cite{minett2011eco, kazemi2017predictive}, and pedestrian safety \cite{11, 12, amin2017implementation} systems are amongst the most important automated driving applications. The first generation of safety applications was designed by virtue of local sensors such as radars and cameras. Local sensors provide a mediocre level of safety due to their limited sensing range and data processing complexity. Moreover, they noticeably underperform in the presence of occluding obstacles. In order to handle these issues, some other sources of information are required to provide more accurate situational awareness within a broader neighboring area. Vehicle-to-Vehicle (V2V) communication has been proposed to remove this barrier through its omnidirectional and non-line of sight connectivity capabilities. Consequently, the performance of safety applications is expected to substantially improve by V2V communications. Currently, the most promising technology under consideration for V2V communication is the Dedicated Short Range Communication (DSRC) \cite{13} technology.

Adaptive Cruise Control (ACC) is one of the most
demanding automated driving applications. In
comparison with its predecessor, i.e. conventional cruise
control which had been solely designed to provide a fluctuation-free driver-specified velocity, ACC is also responsible for
sustaining a certain level of safety by continuously tracking the
vehicle longitudinal distance from its immediate leader and
keeping this distance within a safe range. One step ahead in
cruise technology would result in Cooperative Adaptive Cruise
Control (CACC), which also leverages the V2V communication. This makes it more powerful to
simultaneously preclude collision and maximize traffic
throughput compared to ACC \cite{33}. However, many CACC
challenges still exist which need to be addressed. For instance,
the CACC application should be robust against other vehicles' maneuvers such as unforeseen lane
changes \cite{33}. Detection
and appropriate reaction to these unexpected vehicle maneuvers
are among the most challenging tasks, even in the normal
driving situations and without CACC imposed constraints. These challenging tasks reveal the criticality and complexity of a well-behaved
CACC design for these scenarios. Even though different
theoretical and technical aspects of CACC have been
investigated by researchers \cite{40}, handling interfering vehicles needs more elaborations.

In this paper, we specifically concentrate on cut-in
maneuvers due to their imminent threat, as a vehicle in a stable
CACC platoon has to perform a hard brake reaction when
another vehicle makes a sudden lane change just in front of it. This hard brake reaction is extremely
dangerous and can result in a severe crash \cite{NHTSA2, shladover2015cooperative}. 
Thus, it is well-desired to predict cut-in intention of other drivers
in advance.
Moreover, cutting into the platoon deforms the platoon structure which should be compensated by a proper CACC design. Therefore, a meticulous CACC system should be able to both prevent possible crashes and maintain the normal
platoon formation against entering vehicles from adjacent
lanes.

Based on the above discussion, performance of CACC in these critical driving scenarios is extremely reliant on the accuracy of modeling other drivers’ behavior in the sense of detecting their lane change intentions and predicting cut-in path. This fact requires the introduction of a "lane-change monitoring block", which performs the aforementioned functions, as an inseparable and
essential part of our CACC system.
Specifically, our contributions in this work are listed as follows:
\begin{itemize}
\item A learning-based driver
behavior modeling sub-system is proposed to accomplish an
accurate lane change prediction.
\item A probabilistic framework is designed which employs the results of the lane-change monitoring block and translates it to a cut-in probability
value.
\item A new CACC Stochastic Model Predictive Controller (SMPC) is developed which takes the cut-in probability as its input. This SMPC controller is in charge of adjusting the dynamic
parameters (mainly velocity and spacing error) of the vulnerable vehicles inside the platoon. 
More specifically, it minimizes the spacing
error (deviations from a predefined safe distance) between the
vehicle and its immediate vehicle ahead, while keeping their
velocity difference as close as possible to zero. Concurrently, it responds appropriately to a cut-in maneuver.
\item The overall system architecture is designed and represented as a Time-Triggered Stochastic Impulsive
System (TTSIS) model \cite{16}, originated from the emerging stochastic hybrid systems
(SHS) methodology \cite{14, 15, 16}.
\end{itemize}
To the best of our knowledge, this is the first
cut-in resistant CACC-SMPC design based on a real-time cut-in probability calculation in the literature. 

The rest of this paper is organized as follows. Section \ref{sec:related} is
devoted to related works on proposed driver behavior modeling
methods in the literature. The overall system description is explained in section \ref{sec:system} in which sections \ref{subsec:sys1}, \ref{subsec:sys2} and \ref{subsec:sys3}
state the details of our learning-based cut-in monitoring block,
proposed cut-in probability calculation approach based on that,
and the proposed SHS-based stochastic CACC MPC
framework, respectively. The overall system performance is
evaluated in section \ref{sec:eval}. Finally, section \ref{sec:conc} concludes the paper.

\section{Related Work}\label{sec:related}
Driver is the main source of system
stochasticity in most of the ground ITS frameworks. Each maneuver of a vehicle is an immediate and
direct consequence of its driver's intention, which is applied by a specific set of mechanisms, such
as steering wheel, pedals, and handles \cite{28}. The utilization of these
tools can be directly measured through Controller Area Network (CAN).
However, it is not possible to deterministically
assign a maneuver to a specific pattern of these parameters as
different maneuvers may have partially similar sections \cite{31}.
Therefore, a reliable approach is required to
discriminate different driving maneuvers based on measured
patterns of their parameters. The output of this stage could then be
utilized to design an application-specific controller. This
controller would obviously perform smarter compared to the
controller which only acts based on the previous measurements
without any predictive vision of driving scenario. 
The prevailing methods in the literature for driver behavior
modeling, and some of the important
proposed designs for adaptive cruise controller are mentioned in this section. 

One of the important research mainstreams in driver behavior modeling is based on utilizing classification
methods, such as Support Vector Machine (SVM) and
Neural Network (NN), to differentiate between distinguishable driver
behaviors. The main idea behind another major class of driver behavior modeling schemes in the literature, such as Hidden Markov Models (HMMs) and Dynamic
Bayesian Networks (DBNs), is developing a probabilistic causal framework which tries to
find the next most likely driving maneuvers using available data
sequences from the driving history and then chain these predicted
consecutive maneuvers to construct the most probable future
scenario. 

Authors in \cite{19} developed a hierarchical classifier for
observed scenes of the host vehicle from remote vehicle's lane change. These scenes were then
assigned to the nodes of the hierarchy in the model to specify a
pattern from the top nodes to the leaves. However, their overall scheme is not generalizable to other contexts, such as potential maneuver alternatives, since it is remarkably specialized.

SVM-based methods are proposed to classify lateral actions
of drivers based on detection of preparatory behaviors, vehicle
dynamics, and the environmental data prior to and during the
maneuvers such as lane change \cite{20}. 
A Relevance Vector Machine (RVM), was employed in \cite{21} to distinguish between
lane change and lane keeping maneuvers. In \cite{22}, feed forward artificial neural networks are used to predict
the trajectory of the vehicle based on its movements
history. The goal was to study the possibility of accurate
movement prediction for a lane changing vehicle by an
autonomous driving vehicle.

An Object-Oriented Bayesian Network (OOBN) is utilized
to recognize special highway driving maneuvers, such as lane change \cite{23}. This
approach models different driving maneuvers as vehicle-lane
and vehicle-vehicle relations on four hierarchical levels which
can tolerate uncertainties in both the model and the
measurements. 
A finite set of driving behaviors are classified and future
trajectories of the vehicle are predicted based on currently
understood situational context using a DBN-based model \cite{24}. 

Hidden Markov model (HMM) technique has been widely
utilized to associate the observable time series of the vehicle to
the unobserved driver intentions sequence during his
maneuvers \cite{25, 26, 31, 32, 27, 28, 29, 30}. Some pioneer works in driver behavior
modeling, \cite{25, 26}, proposed a decomposition of driver
behaviors into small scale and large scale categories. Time
sequence of unobserved large scale driver actions are assumed
to have Markovian property and HMM is suggested as an
acceptable method to model this sequence. This modeling
approach accuracy was validated by its results of the lane
change maneuver prediction. Sensory collected information
was used as the observation set in the designed HMM predictor.

Using the data from V2V communication, two HMMs were
utilized to discriminate different types of driver lane change
intent, namely dangerous and normal \cite{31, 32}. A trajectory
prediction stage and an MPC controller were mounted on top of
the lane change prediction algorithm to manage reformation of
a new CACC string after cut-in.

A controller for a CACC string which takes into account both
V2V and non-V2V equipped vehicles was designed in \cite{33}.
This controller tries to handle cut-in and cut-out scenarios with
a smooth reaction to the new condition of the host vehicle’s
lane. No prediction is performed in
this work to detect the cut-in or cut-out scenarios in advance. Another CACC design based on switched sampled-data model is presented in \cite{40} which investigates the stability problem in the presence of sensor failures. 

In our work, a learning-based driver behavior modeling method is combined with an MPC design in a probabilistic manner to improve the overall CACC performance.

\section{System Description} \label{sec:system}
\begin{figure}[t]
\begin{center}
\includegraphics[width=0.95\linewidth]{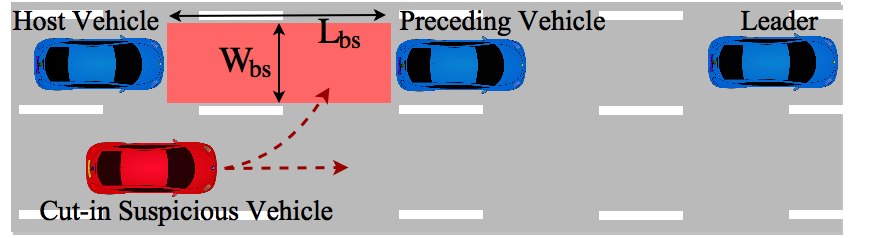}
\end{center}
\caption{Host vehicle, cut-in suspicious vehicle and bad-set}
\label{fig:framework}
\end{figure}

In our framework, which is schematically depicted in Fig.
\ref{fig:framework}, the vehicle inside the platoon, which is directly affected by
the cut-in suspicious vehicle, is referred to as the host vehicle.
The immediate vehicle in front of the host vehicle is known as
the preceding vehicle, and the first vehicle of the platoon is the
leading vehicle or leader. The dangerous area in front of the
host vehicle is referred to as bad-set. This area and its
dimensions will be discussed in details later.

Although, detection of cut-in by the vehicle itself is
beneficial to some applications such as lane keep assist system
(LKAS) and blind spot warning (BSW), CACC and platooning
need the lane change maneuver to be detected remotely by the
host vehicle. The remote lane change
detection is required because the host vehicle should react in a
timely manner to avoid hazardous situations.

V2V communication periodically provides the parameters of
the cut-in suspicious vehicles via broadcasting basic safety
messages (BSM) \cite{13, 34}. In our model, we assume that the
host vehicle, which is in a stable condition in the platoon,
periodically receives the BSMs of its surrounding vehicles and
continuously traces them prior to any probable cut-in maneuver.
From BSM part one of the SAE J2735 standard, \cite{34}, we utilize
the following parameters for our behavior modeling: latitude,
longitude, elevation, speed, heading, steering wheel angle, 4-way acceleration set, and the vehicle size. The latitude, longitude,
and elevation represent the location of the vehicle’s center of
gravity in the WGS-84 coordinate system. The 4-way
acceleration set consists of acceleration values in 3 orthogonal
directions plus yaw rate, which are calculated based on the
assumption that the front of the vehicle is toward the positive longitudinal axis, right side of it is the positive lateral axis, and
clockwise rotation as the positive yaw rate.

In CACC platooning, a safe longitudinal gap must be
continuously kept between every two consecutive vehicles. The
deviation from the safe gap, which is known as spacing error,
should remain as small as possible to reduce the risk of collision
and take the advantages of platoon formation, such as lower fuel
consumption and higher traffic throughput \cite{35}. As mentioned,
we define \textit{bad-set} as the dangerous area in front of the vehicle
in which the safe gap is violated. In other words, our bad-set is
a rectangle aligned to the road surface in front of the host
vehicle, while its longitudinal dimension, $L_{bs}$, depends on the
platoon speed and is equal to the desired longitudinal safe gap
and its lateral dimension, $W_{bs}$, is the lane width. The front bumper of the host vehicle is always
located at the center of the bad-set rear lateral edge. These
definitions are illustrated in Fig. \ref{fig:framework}.

The goal of our “lane-change monitoring block” is tracking and
predicting the trajectory of all of the vehicles in the adjacent lanes
of the host vehicle. The model should not only predict the
immediate kinematics of the vehicles, but also the high-level driving maneuvers. Therefore, the position of
neighboring vehicles should be predicted for multiple future
steps based on their current and previous communicated
information. The number of required prediction steps is
determined by the duration of a complete high-level maneuver
and denoted by $S_m$. This multi-step prediction is then used to
determine the probability of unsafe lane change which is passed
to the SMPC for better estimation of the required inter-vehicle spacing gap.
\subsection{Lane Change Monitoring Block Design}\label{subsec:sys1}
Each lane change maneuver consists of four separate phases:
Intention phase, Preparation phase, Transition phase, and the
Completion phase \cite{36, 37}. It is worth mentioning that some more complicated maneuvers, such as overtaking, have also been investigated in the literature. For instance two-phase and five-phase overtake modeling frameworks are proposed in \cite{barmpounakis2017decision}  and \cite{wilson1982driving}, respectively.  
The lateral acceleration
and lateral speed in a lane change maneuver are bounded by the comfortable lateral
acceleration threshold and the maximum tolerable lateral speed,
respectively \cite{38}. To safeguard a smooth transition of the
vehicle between lanes, the acceleration is bounded by -0.2g and
0.2g \cite{39}. 
Our model is designed to not only predict the
immediate kinematics of the vehicles in the transition phase but
also the complete four-phase lane change maneuvers.
The trajectory of each remote vehicle is modeled as a time series.
In our model, we separate the learning
of lateral and longitudinal behaviors of the driver as they are
influenced by different control inputs. 

Artificial neural networks (ANNs) are one of the most famous
tools for description and prediction of nonlinear
systems \cite{khosroshahi2016surround, connor1994recurrent}. Neural networks with hidden units can principally
predict any well-behaved function. In the case of time series, in order to handle the
dependency of the prediction to a finite set of past values and
time varying nature of the input signals, neural network
topologies need to be equipped with a short term memory
mechanism which is called the feedback delay. In this work, we used feedback delay-
based ANNs, namely nonlinear autoregressive (NAR), nonlinear autoregressive
exogenous (NARX) and recurrent neural networks (RNN)
toward driver behavior and lane change prediction.

NARX is a neural network with feedback
delay that can be trained and used to predict a time series from
its past values and an exogenous one, in spite of NAR which does not rely on any external inputs. 
We use NAR model
to predict the future pattern of different system inputs, i.e. steering wheel angle,
yaw rate, heading, speed, and longitudinal acceleration, based on their currently available values. 
A NARX model is employed to
predict the longitudinal trajectory of the vehicle during the lane
change using some of the previously estimated sequences of
input signals as the exogenous input. 
The exogenous inputs in our framework are yaw rate, heading, speed, and longitudinal
acceleration. 

Finally, an RNN is adopted to model the lateral trajectory of
the vehicle based on the predicted input signals. 
RNNs can use their internal memory to process
arbitrary sequences of inputs. The input signals to our lateral
position prediction RNN are steering wheel angle, yaw rate, and
heading. Using the internal memory, the RNN can distinguish
between different maneuvers with partially similar input
signals. For example, a steering due to the road curvature might
look partially similar to the one from lane change maneuver, but
the RNN can learn to distinguish between these two maneuvers by
looking at a longer history of the signals or other input signals, such as road curvature. In the former case, the RNN should also be trained on other maneuvers which share the same input signal patterns in a portion of their lifetime.

All of the ANN models are batch trained and the training
phase is offline due to the low computational cost of batch
training and insufficient accessible data for online training. 
In order to use the full capability of neural networks, the input signals for all
ANNs are normalized to [-1, 1] range. Then, the input signals are differenced
to remove the linearity and improve the nonlinearity prediction process. The
resulting time series is known as integrated time series.
The value of predicted location can be reconstructed by adding the first actual value to
the estimated difference in the series. 

\begin{figure}[t]
\begin{center}
\includegraphics[width=0.65\linewidth]{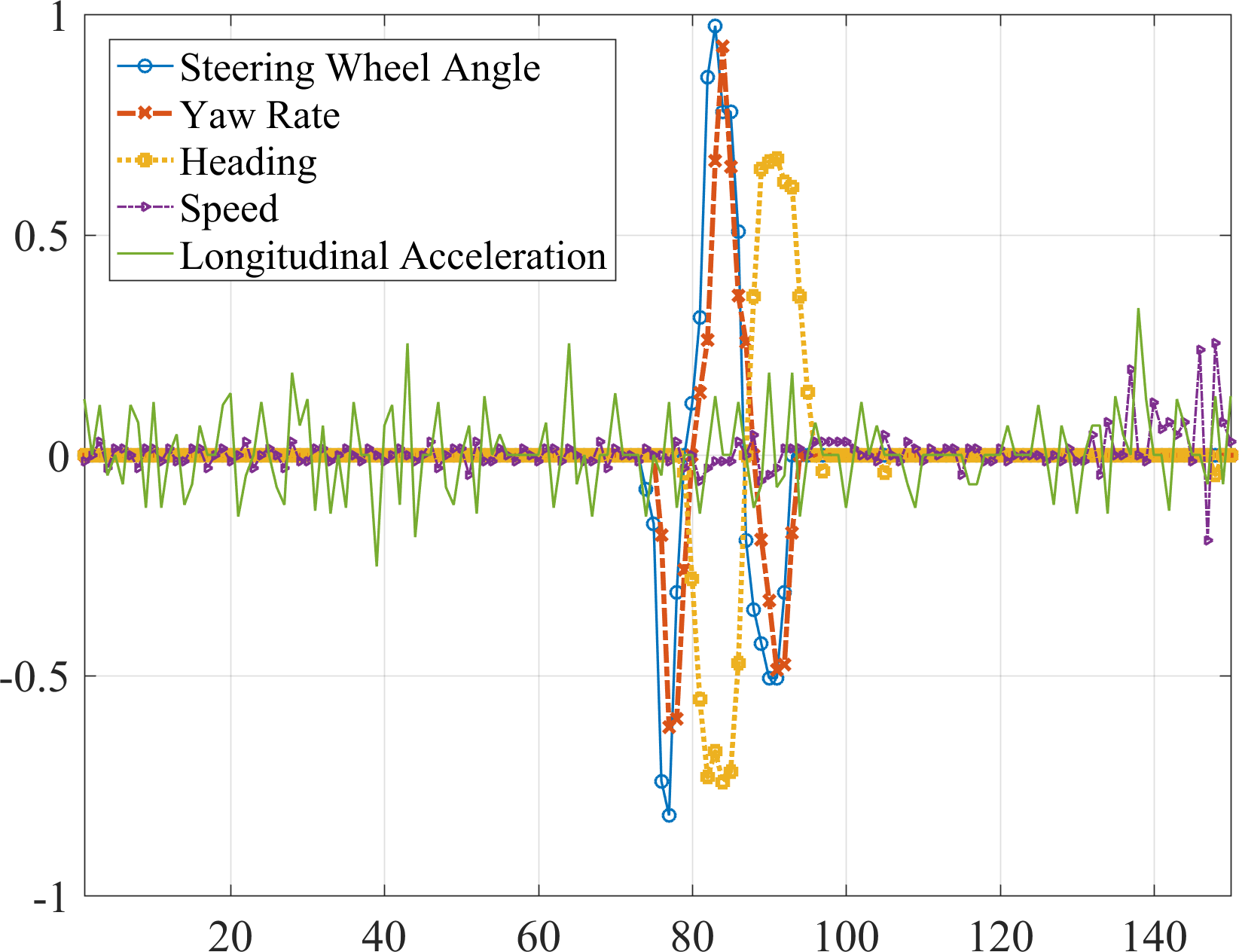}
\end{center}
\caption{Smoothed, Normalized, and Integrated input signals of a single lane change maneuver}
\label{fig:signals}
\end{figure}

To mitigate the effect of noise, small
variations of input signals, based on the nature of the signal, are
filtered to smooth the time series and mitigate the effect of
noise. Variation smaller than 3 degrees, 0.1 $rad$, 0.1 $m/s$, and
0.1 $m/s^2$ are removed from steering wheel angle, heading,
speed, and longitudinal acceleration, respectively. The resulting
input signals during one maneuver is shown in Fig. \ref{fig:signals}. 

All of our ANNs have a hidden layer with 20 nodes and 15
step short term memory, which means that they are using the
past information of 1.5 seconds for future prediction. The
required prediction steps are also set to 10 steps for all of the
ANNs, $S_m = 10$, which means that the we are predicting the
behavior of the driver for 1 second in the future, since the driver
can change his decision and behavior beyond this time \cite{43}.
As mentioned
before, the NAR is used to model the patterns of input signals
to the system. The NARX and RNN
are used to model and predict the longitudinal and lateral
position of the vehicle, respectively. The longitudinal position
of the vehicle is modeled based on the predicted values of
heading, speed, and longitudinal acceleration as external inputs.
On the other hand, the RNN should not
only predict the future lateral position of the vehicle, but should
also distinguish between different lateral maneuvers. Therefore,
the lateral model is also trained with some curve road data to be
able to differentiate between different lateral movements.

\begin{figure}[t]
\begin{center}
\includegraphics[width=0.99\linewidth]{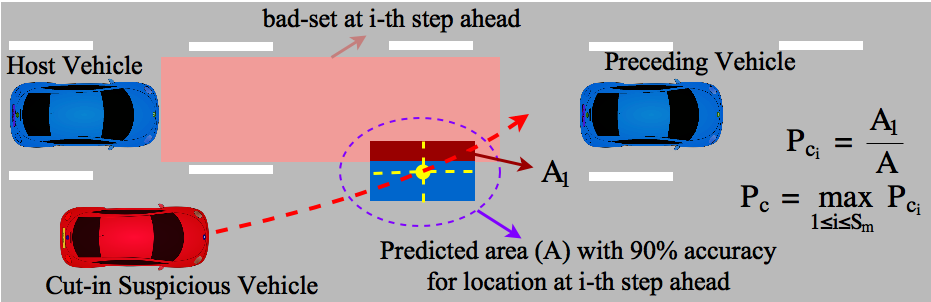}
\end{center}
\caption{Procedure of cut-in probability ($P_c$) calculation}
\label{fig:probability}
\end{figure}
\subsection{Cut-in Probability Calculation}\label{subsec:sys2}

The results of the proposed cut-in prediction scheme is now
applied to find a single value between 0 and 1 which represents
the overall cut-in probability. This probability, which is denoted
by $P_c$ from now on, will be fed to our SMPC as its input. SMPC
design details are discussed in the following subsection. 
At each prediction 
cycle we have $S_m$ predicted future values for each of the longitudinal and lateral relative positions of the suspicious cut-in vehicle. In our implementation, each of these $2 \times S_m$ predicted values comes with a specific 90 percent confidence level. Hence, we have $S_m$ rectangular areas, each of them determines the predicted area for the position of the cut-in vehicle in the corresponding upcoming time step with 90 percent accuracy. We take the most conservative approach to define the cut-in probability, $P_c$ as follows:
\begin{itemize}
\item Each of these $S_m$ rectangles, ($A$ in Fig. \ref{fig:probability}), is intersected with the host vehicle's bad-set at that moment
and its intersection area, ($A_1$ in Fig. \ref{fig:probability}), is calculated.
\item The resultant intersection area, ($A_1$), is
normalized by dividing it by the corresponding predicted area
value, ($A$), to calculate the probability
value of being inside the bad-set for each of these predictions.
\item The maximum
value amongst these $S_m$ probabilities is selected as the $P_c$ value
for that prediction cycle.
\end{itemize}
For more clarification, this procedure
for the $i^{th}$ step prediction, ($1 \leq i \leq S_m$), is depicted in Fig. \ref{fig:probability}. In this paper, one second ahead prediction is targeted which
is equivalent to $S_m = 10$, due to the DSRC baseline
information broadcasting frequency (10 Hz). It is worth mentioning that this frequency could be easily supported by most of the currently available commercial GPSs like what is used in this work's dataset \cite{42}.

\subsection{CACC-Stochastic Model Predictive Controller Design}\label{subsec:sys3}
Considering a CACC platoon of vehicles, the spacing error
of the $i^{th}$ following vehicle is defined as follows \cite{39}:
\begin{align}
\delta_i = x_{i-1}-x_i-hv_i-L_i-d_0
\end{align}
for all $i \in \{1,2, \dots, n\}$, where $x_i$ and $v_i$ are longitudinal position
and velocity of the $i^{th}$ following vehicle, respectively ($x_0$ stands for the longitudinal position of the lead vehicle); $h$ headway which introduces a speed dependent spacing policy in addition to $d_0$ which is a constant minimum desired distance between each vehicle and its preceding vehicle in the platoon, and $L_i$ is the length of the $i^{th}$ vehicle. 
Based on these definitions, longitudinal dimension of the bad-set, $L_{bs}$, for $i^{th}$ vehicle could be represented as:
\begin{align}
L_{bs}=hv_i+d_0
\end{align}
Then, the linearized dynamics of the $i^{th}$ following vehicle is modeled as follows \cite{39}:
\begin{align} \label{dynamics1}
\dot \delta_i &= v_{i-1}-v_i-h \dot v_i \\
\Delta \dot v_i &= a_{i-1}-a_i \\
\dot a_i &= -\dfrac{a_i}{\zeta_i}+\dfrac{u_i}{\zeta_i} \label{dynamics3}
\end{align}
where $\zeta_i$ is the engine time constant, $a_i$ is the acceleration of the $i^{th}$ vehicle, and $u_i$ is an input signal which comes from an MPC controller. However, due to the communication delay each vehicle receives the delayed version of its preceding vehicle's acceleration value. Denoting the communicated acceleration of $i^{th}$ vehicle at receivers by $\bar a_i(t)$, the state space equation of a vehicle in a CACC system could be represented as follows
\begin{align} \label{ss_dynamics}
\dot x_i(t) &= A_i x_i (t) + B_i u_i(t) + G_i \bar a_{i-1}(t)
\end{align}
with state vector $x=[\delta_i \quad \Delta v_i \quad a_i]^T$, and
\begin{align} \label{ss_params}
A_i=\begin{bmatrix}0 & 1 & -h \\ 
0 & 0 & -1 \\
0 & 0 & -\frac{1}{\zeta_i}\end{bmatrix} \enskip 
B_i=\begin{bmatrix}
0 \\ 0 \\ -\frac{1}{\zeta_i}
\end{bmatrix} \enskip
G_i=\begin{bmatrix}
0 \\ 1 \\ 0
\end{bmatrix}
\end{align}
However, delay of the communication network is not considered in this work, so $\bar a_{i-1}(t) = a_{i-1}(t)$. 

An MPC controller with three primary objectives is required to control the platoon system described by (\ref{ss_dynamics}). The controller must compute the input signal $u_i$ to minimize the spacing error, keep the velocity of the host vehicle as close as possible to its preceding vehicle velocity, and finally, respond appropriately to a cut-in vehicle based on our prediction of the driver
behavior. To this end, the system dynamics (\ref{ss_dynamics}) is discretized
and an optimal control problem, which satisfies the
aforementioned control goals, is defined. The continuous time
dynamics of the system is discretized using the Euler forward
method with a time step $T_s$:
\begin{align} \label{ss_dynamics_dis}
\dot x_i[k+1] &= A_i^k x_i [k] + B_i^k u_i[k] + G_i^k \bar a_{i-1}[k]
\end{align}
where
\begin{align} \label{ss_params_dis}
A_i^k=\begin{bmatrix}1 & T_s & -hT_s \\ 
0 & 1 & -T_s \\
0 & 0 & 1-\frac{T_s}{\zeta_i}\end{bmatrix} \enskip 
B_i^k=\begin{bmatrix}
0 \\ 0 \\ -\frac{T_s}{\zeta_i}
\end{bmatrix} \enskip
G_i^k=\begin{bmatrix}
0 \\ T_s \\ 0
\end{bmatrix}
\end{align}
Hereinafter, for simplicity of notation, we use $A, B, G, x$, and $u$ instead of $A_i^k, B_i^k, G_i^k, x_i$, and $u_i$, respectively. The cost function of the optimal control problem is defined based on the primary objectives of the controller:
\begin{align} \label{cost_1}
&J[k]=\sum_{i=0}^{N-1}c_\delta \delta^2[k+i] + c_v \Delta v^2[k+i] + c_u \Delta u^2[k+i]\\
& = \sum_{i=0}^{N-1}[x^T[k+i|k]Qx[k+i|k]+u^T[k+i|k]Ru[k+i|k]]\nonumber
\end{align}
where $N$ is the control horizon, $c_\delta$, $c_v$, and $c_u$ are weighting coefficients reflecting the relative importance of each term and
\begin{align} \label{delta_u}
\Delta u[k+n]= u[k+n] - u[k+n-1]
\end{align}
which is added to the cost function as an extra term to bound the jerk and prevent fast variations of the input signal. This constraint could be interpreted as comfort ride. The MPC law finds the optimal input sequence $\bm u^*[k]$ which minimizes the predicted cost function (\ref{cost_1}) at each time instant:
\begin{align} \label{opt_problem}
&\bm u^*[k] = \arg \min_u J[k] \\ \nonumber &\text{subject to} \quad \left\{\begin{matrix}
x_{min} \leq x[k+i|k] \leq x_{max} \\
u_{min} \leq u[k+i|k] \leq u_{max}
\end{matrix} \quad i =1, \dots, N-1 \right.
\end{align}
To solve this MPC problem, the future values of the preceding vehicle's acceleration are required. These values are obtained from the aforementioned NAR neural network.
\subsubsection{Conventional MPC Design}
In this section, MPC design problem without incorporating
the calculated cut-in probability, $P_c$, is investigated. We referred
to this MPC design as conventional design in this paper. The values of $a_{i-1}[k]$ could be considered as a measured disturbance when its model is available to the MPC controller. Then, the system equations could be rewritten in the standard form as
\begin{equation} \label{system}
\bar {\bm x}[k+1]=\bar A\bar {\bm x}[k]+\bar Bu[k]
\end{equation}
where $\bar {\bm x}[k] = \big[\bm x[k],\bm v[k]\big]^T$ is the augmented state vector and the measured disturbance state vector $\bm v[k] = \big[v_0, v_1, \dots v_{N-1}\big]$ is defined as
\begin{align} \nonumber
\begin{matrix}
v_0[k+1] = v_1[k] \\
v_1[k+1] = v_2[k] \\
\vdots \\
v_{N-2}[k+1] = v_{N-1}[k] \\
v_{N-1}[k+1] = v_{N-1}[k] 
\end{matrix} \qquad  \enskip \begin{cases}
v_0[0] =a_{i-1}[0] \\
v_1[0] = a_{i-1}[1] \\
\vdots \\
v_{N-2}[0] = a_{i-1}[N-2] \\
v_{N-1}[0] = a_{i-1}[N-1] 
\end{cases}
\end{align}

In the receding horizon implementation of the MPC problem (\ref{opt_problem}), only the first element of the optimal input sequence $u^*[k]$ is selected as the input to the system and the whole process is repeated at each time step. However, designing a receding horizon controller based on a finite-horizon cost function does not guarantee the stability and optimality of the closed loop system \cite{41}. This problem can be avoided by defining an infinite prediction horizon for the cost function:
\begin{align} %\label{infinite_cost}
J[k] &= \sum_{i=0}^{\infty}[x^T[k+i|k]Qx[k+i|k]+u^T[k+i|k]Ru[k+i|k]] %\nonumber
\end{align}
However, to have finite number of variables in the MPC optimization problem, a dual-mode prediction approach can be utilized in which the predicted input sequence is defined as
\begin{align} \label{control_law}
u[k+i|k]=\begin{cases}
{\bm u}^*[k+i|k] & i = 0, 1, ..., N-1\\
Kx[k+i|k] & i = N,N+1,...
\end{cases}
\end{align}
Then, by choosing a terminal weighting matrix, denoted by $\bar Q$, in a way that $x^T[k+N|k] \bar Q x[k+N|k]$ is equal to the cost over the second mode of the predicted input sequence, the infinite cost $J$ can be rewritten as %(\ref{infinite_cost}) can be rewritten as
\begin{align} \label{finite_cost}
J &[k] = \sum_{i=0}^{N-1}[x^T[k+i|k]Qx[k+i|k]\\ \nonumber
& +u^T[k+i|k]Ru[k+i|k]] + x^T[k+N|k] \bar Q x[k+N|k]
\end{align}

\begin{theorem}
\textbf{(Stability) } The state variables of system (\ref{system}), $x[k]$, asymptotically converge to zero, i.e. the system is asymptotically stable, under the control law (\ref{control_law}) if predicted cost $J[k]$ is an infinite cost, $(A, Q^{1\/2})$ is observable, and the tail $\tilde{\bm u}[k]$ is feasible for all $k>0$ where
\begin{equation}
\tilde{\bm u}[k+1]=\big[\bm u^* [k+1|k], \dots , K\bm x^*[k+N|k]\big]
\end{equation}
\end{theorem}
Therefore, selecting $Q$ and $\bar Q$ which satisfy the first two
conditions, the stability and convergence of the closed loop
system rely on the assumption that tail $\tilde{\bm u}[k]$ is feasible for all $k>0$.   To this end, a set of extra constraints on the state vector should be satisfied at each time instant $k$.
\begin{theorem}
	\textbf{(Recursive feasibility) } The MPC optimization (\ref{opt_problem}) with the cost function $J[k]$ defined in (\ref{finite_cost}) is guaranteed to be feasible at all time $k>0$ if a new constraint $\bm x[k+N|k] \in \Omega$ is met, provided it is feasible at $k=0$ and terminal constraint set $\Omega$ satisfies
	\begin{itemize}
		\item The following constraints are satisfied for all points in $\Omega$, ( i.e. $\bm x[k+N|k] \in \Omega$)
		\begin{equation} 
			 \begin{cases}
			u_{min} \le K\bm x[k+N|k] \leq u_{max} \\
			\bm x_{min} \le \bm x[k+N|k] \leq \bm x_{max}
			\end{cases}
		\end{equation}
		\item $\Omega$ is invariant in the second mode of (\ref{control_law}) which means
		\begin{equation} 
			\bm x[k+N|k] \in \Omega \quad \Rightarrow \quad \bm (A+BK)x[k+N|k] \in \Omega
		\end{equation}
	\end{itemize}
\end{theorem}
It is shown that the largest possible $\Omega$ is derived by
\begin{align}
	\Omega = \{\bm x: \quad &u_{min} \leq K(A+BK)^i \bm x \leq u_{max}, \\ \nonumber &\bm x_{min} \leq (A+BK)^i \bm x \leq \bm x_{max}, \quad i=0,1,\dots\}
\end{align}
To show that $\Omega$ is invariant over the infinite horizon of the second mode of (\ref{control_law}), constraint satisfaction should be checked over a long enough finite horizon ($N_c$). Here, $N_c$ is the smallest number which satisfies the following equations
\begin{align} 
	\begin{cases}
	\underbar{$u$} = \min_x K(A+BK)^{N_c+1} \bm x \\ 
	\overline{u} = \max_x K(A+BK)^{N_c+1} \bm x 	
	\end{cases}
\end{align}
such that
\begin{align}
	\begin{cases} 
	u_{min} \leq K(A+BK)^i \bm x \leq u_{max} \quad
	i = 0, \dots, N_c\\
	u_{min} \leq \underbar{$u$} \leq u_{max}
	\end{cases}
\end{align}
Having $N_c$ found, adding the following constraints to the MPC problem guarantees the feasibility of the controller:
\begin{align} 
	&u_{min} \leq K(A+BK)^i \bm x[k+N|k] \leq u_{max}, \\ \nonumber
	&x_{min} \leq (A+BK)^i \bm x[k+N|k] \leq x_{max}, \\ \nonumber & \qqquad i = 0, 1, \dots, N_c
\end{align}

\subsubsection{MPC design: Incorporating cut-in probability}
The designed MPC controller in the previous section satisfies
our first two primary goals, namely spacing error and velocity
error minimization. However, the controller should be able to
react appropriately if a cut-in suspicious vehicle enters the
platoon unexpectedly and pushes the host vehicle to decelerate
to reestablish the safe distance.

Heretofore, the probability of the suspicious vehicle's cut-in
trajectory intersection with the host vehicle's bad-set has been
determined. Based on this, we propose a new stochastic
definition for the spacing error:
\begin{align} \label{spaceing_error_2}
\delta_i = \dfrac{x_{i-1}-x_i}{2-e^{-\alpha P_c}}-hv_i-L_i-d_0
\end{align}
where $P_c$ is the probability of the cut-in vehicle being in the bad-
set of the host vehicle, and is a constant control parameter
which adjusts the reaction sensitivity of the MPC controller to the cut-in probability. Clearly, when the probability is one,
assuming $\alpha$ has been set to a sufficiently large number, the
controller starts doubling the distance from its current
preceding vehicle by halving the enumerator of the first term
in the proposed equation for spacing error, (\ref{spaceing_error_2}). Consequently,
the cut-in vehicle has enough safe gap to enter the CACC
platoon. On the contrary, when the probability is zero, the
suspicious vehicle is not expected to cut in or it has the safe
distance from the host vehicle for its maneuver. This zero
probability sets the denominator of (\ref{spaceing_error_2}) to one, which means the host vehicle keeps the normal safe distance, $hv_i+d_0$, from its preceding vehicle.

Although the stability and feasibility of the controller is
already guaranteed for the MPC design with deterministic
spacing error, it should be proved under the new circumstances
due to the stochastic spacing error definition. 

\begin{figure}[t]
\begin{center}
\includegraphics[width=0.5\linewidth]{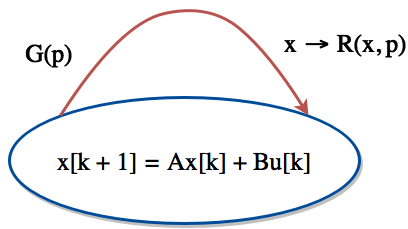}
\end{center}
\caption{A hybrid model for the system incorporating the cut in probability}
\label{hybrid_automata}
\end{figure}

To this end, utilizing a Stochastic Hybrid System (SHS)
design is beneficial. A Time-Triggered Stochastic Impulsive
System (TTSIS) SHS model, \cite{16}, incorporating the probability
of an upcoming cut, is utilized to this end as shown in Fig. (\ref{hybrid_automata}).

In this model, $p\in[0,1]$ is a random variable which represents the cut-in probability, $G(p)$ is a guard condition that
must be hold for the discrete transition (time trigger), and $R(x,p)$ is a reset function which describes the changes in the
continuous states after the transition. The stochastic nature of
our system comes from the dependency of the guard on the
random variable $p$. 
\begin{prepos}
	The TTSIS of Fig (\ref{hybrid_automata}) is stable under the designed MPC controller if the following condition is satisfied:
	\begin{itemize}
		\item There is a finite invariant region $\mathpzc{S}_R$ in which
			\begin{equation}
				\begin{matrix}
				    x[k] \in \mathpzc{S}_R \quad \rightarrow \quad R(x[k],p) \in \mathpzc{S}_R \\
					\forall p \in [0,1]. \quad
					x_{min} \leq x[k] \leq x_{max}
				\end{matrix}
			\end{equation}
	\end{itemize}
\end{prepos}

The region $\mathpzc{S}_R$ is a subset of the region of attraction for the MPC law, denoted by $\mathpzc{S}_\Omega$ ($\mathpzc{S}_R \subset \mathpzc{S}_\Omega$). Here, $\mathpzc{S}_\Omega$ is defined as the set of all initial states from which a sequence of inputs exists
that forces the state predictions to reach the terminal constraint
set Ω in the first mode of control law (\ref{control_law}), i.e.
\begin{equation}
	\mathpzc{S}_\Omega = 
	\left\{
		\begin{aligned}
			& x[0] : \quad \exists \bm u[0], \quad \bm x[N|0] \in \Omega,\\ 
			&\text{s.t.} \quad
 \begin{cases}
				x_{min} \leq x[i|0] \leq x_{max}, \quad \forall i \leq N \\
				u_{min} \leq u[i|0] \leq u_{max}, \quad \forall i \leq N-1
			\end{cases}
		\end{aligned} 
	\right.
\end{equation}

Therefore, to guarantee the stability of the system over a
larger set of conditions, the constraints on the system states
should be relieved as much as the safety is not violated. To this end, we set the constraint on the spacing error to
$[\delta_{min}, \delta_{max}] \in (-hv_i-L_i-d_0+\delta_s, +\infty)$ for the case of no cut-in detected, where $\delta_s$ is the minimum desired safe gap between the vehicles. Clearly, after each discrete transition, the value of spacing error can only jump with a value between $-hv_i-L_i-d_0$ and $hv_i+L_i+d_0$. However, for the case of a positive cut-in probability,
we should choose a more conservative constraint, $[\delta_{min}, \delta_{max}] \in (-hv_i-L_i-d_0 + (x_{i-1}-x_{rv})+\delta_s, +\infty)$ to assure the collision avoidance where $x_{rv}$ is the position of the cut-in suspicious vehicle. Finally, if the is no constraint found
which can guarantee the feasibility, the driving situation is considered as an unsafe or a harsh maneuver and the controller
temporarily is overwritten with $u_i$ set to the maximum possible
deceleration (usually up to $-10 m/s^2$ \cite{gietelink2006development}, to prevent the
collision) or the maximum pre-defined acceleration of the
vehicle till the feasibility of the controller can be guaranteed
again.

\section{Evaluation} \label{sec:eval}
In this section overall performance of the proposed system
framework is evaluated using realistic driving scenarios from
Safety Pilot Model Deployment (SPMD) dataset \cite{42}. First, the
practicality of our cut-in trajectory prediction method is shown
by its performance comparison versus the kinematic-based
trajectory prediction as a ground truth. Next, noticeable better
behavior of the proposed SMPC controller versus the
conventional MPC is discussed.

\subsection{Cut-in Trajectory Prediction Performance Evaluation}
To evaluate the performance of our method, we extracted 90
lane change maneuvers from the BSMs generated by
participating vehicles in SPMD dataset in Ann Arbor,
Michigan. BSM broadcast rate had been set to 10Hz in this dataset. Therefore, we have all of the time series recorded in this rate. 
The signals from all 90 maneuvers are concatenated to
create a long univariate time series for each input signal. Finally, a 70-15-15 percent training, cross-validation, and
testing data selection is used for ANNs training and
performance evaluation.

\begin{figure}
\centering
\begin{tabular}{cc}
\subfigure[]{\includegraphics[width=.22\textwidth]{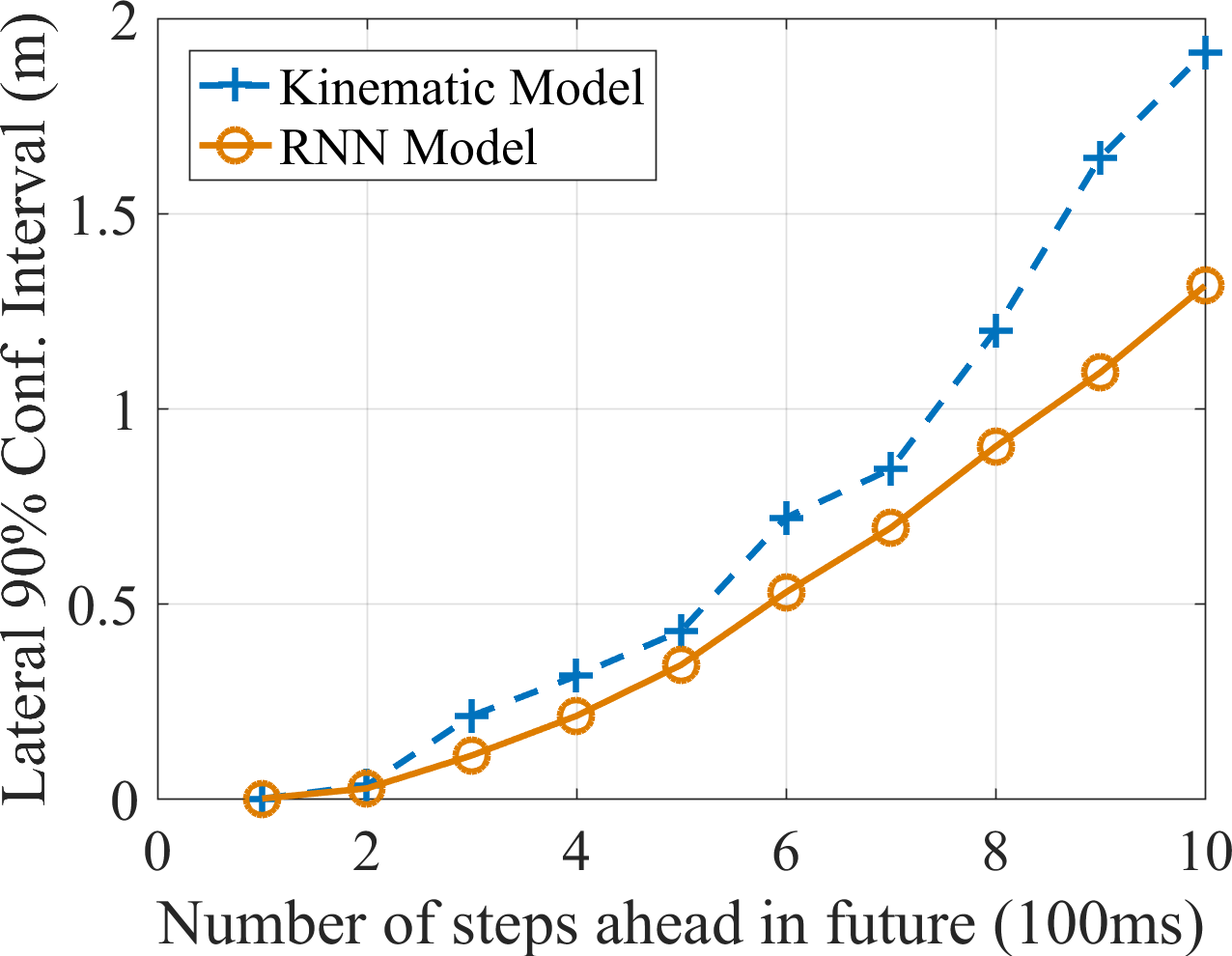}\label{fig:rnn1}} & 
\subfigure[]{\includegraphics[width=.22\textwidth]{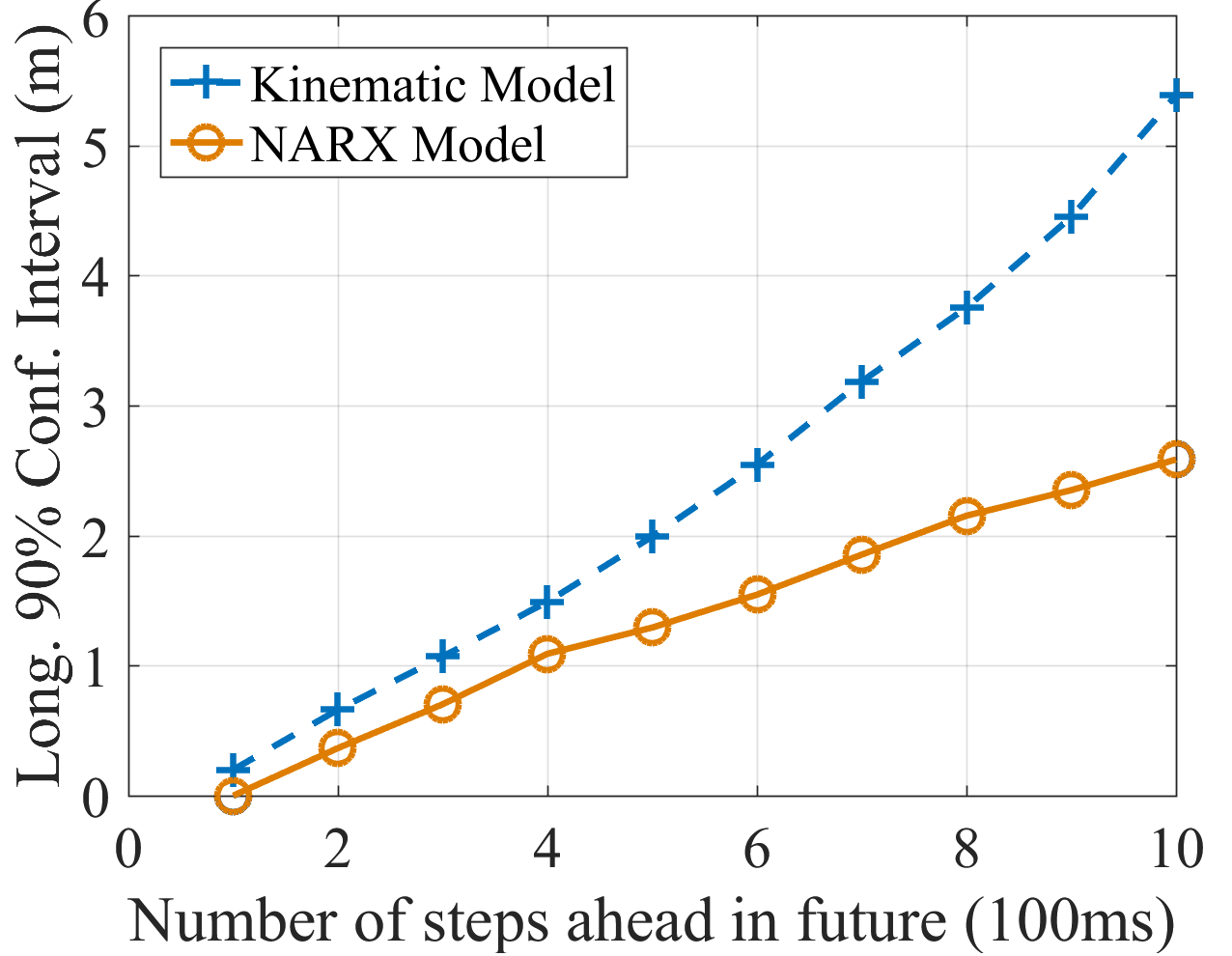}\label{fig:rnn2}} \\
\end{tabular}
\caption{Comparison of 90-percentile conf. interval of the Kinematic and RNN models for different prediction steps. (a) Lateral and (b) Longitudinal predictions}
\end{figure}

\begin{figure}
\centering
\begin{tabular}{cc}
\subfigure[]{\includegraphics[width=.23\textwidth]{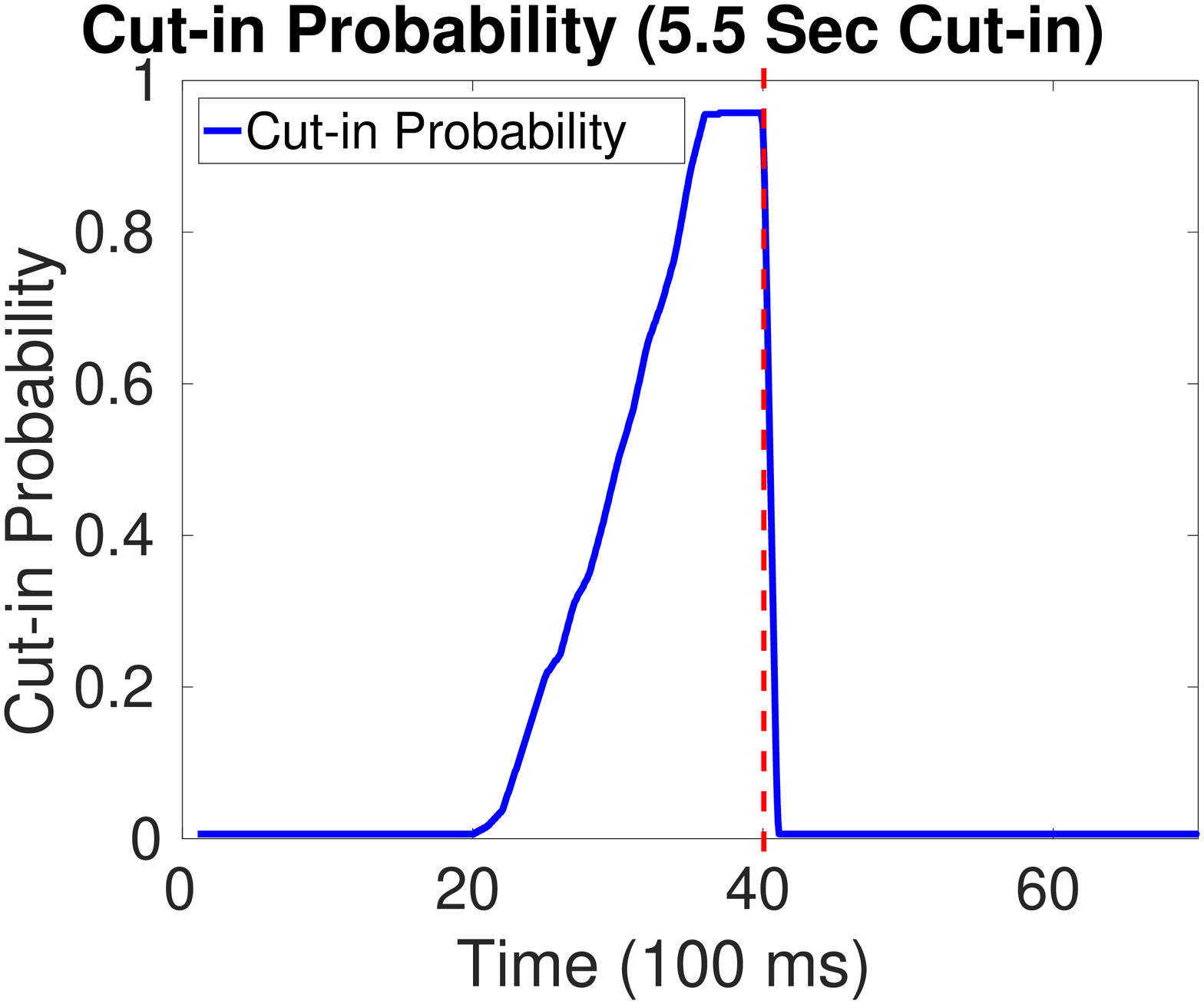}\label{fig:p5}} & 
\subfigure[]{\includegraphics[width=.23\textwidth]{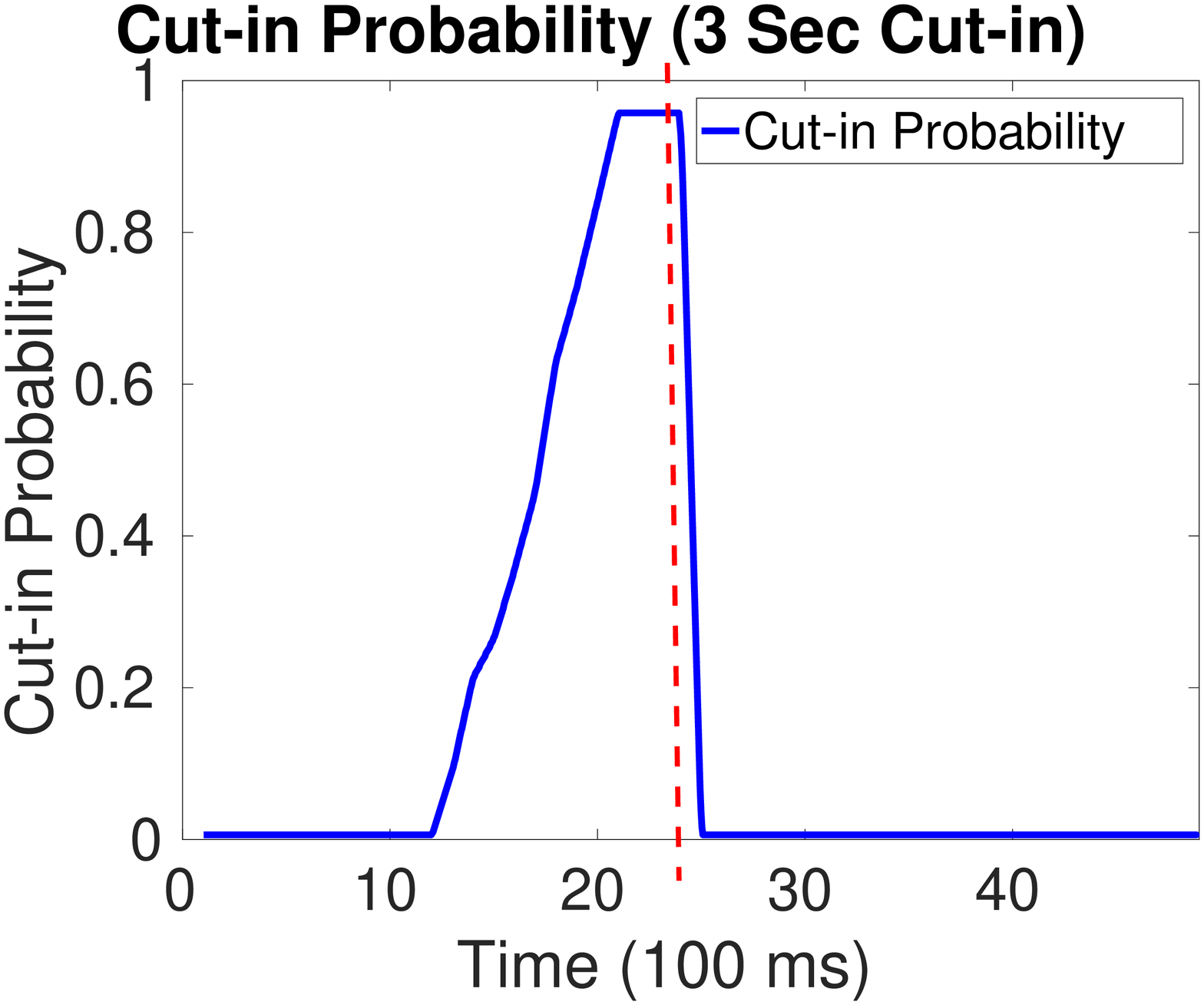}\label{fig:p3}} \\
\end{tabular}
\caption{Cut-in Probability, $P_c$, for the proposed SMPC controller (a) an average 5.5-sec maneuver and (b) a harsh 3-sec maneuver }
\label{fig:prob}
\end{figure}

The performance of two trajectory prediction methods,
vehicle kinematic model, and our trained RNN, for a lane
changing vehicle in terms of lateral confidence levels at each
time step ahead, averaged on all 90 scenarios, are shown in
Fig.~\ref{fig:rnn1}. The classic car model could be represented by
kinematic-based differential equations as follows:
\begin{equation}
\dot x_i = v_i \cos \theta_i, \qquad
\dot y_i = v_i \sin \theta_i, \qquad
\dot \theta_i = \dfrac{v_i}{L_i} \tan \phi_i
\end{equation}
where $x_i, y_i$, and $v_i$ are longitudinal position, lateral position, and velocity of the $i^{th}$ vehicle, respectively. Also, $\phi_i$ denotes the steering angle, $\theta_i$ stands for the angel between the vehicle's instantaneous heading and the road direction, and $L_i = 5$ is the length of the vehicle \cite{44}.

Fig.~\ref{fig:rnn2} shows the same comparison for the longitudinal position prediction.

\subsection{Designed SMPC Performance Evaluation}
In this section, superiority of designed SMPC versus
conventional MPC is investigated. To this end, reactions of
these two different designs to real cut-in maneuvers should be
compared. A general cut-in maneuver duration is between
3.5-6.5 seconds for urban scenarios with the mean of 5
seconds and 3.5-8.5 seconds for highway scenarios with the
mean of 5.8 seconds \cite{45}. For the sake of comparison fairness, two types of cut-in maneuvers, i.e. the harshest type with 3.5
seconds duration, and the average class with 5.5 seconds
duration, have been selected and the outputs of two
aforementioned controllers are compared in each case. In both
cases, the platoon velocity is assumed 27 $m/s$ or 60 $mph$.

Cut-in probabilities, calculated based on the discussed
method in section \ref{subsec:sys2}, for average and harsh maneuvers are
depicted in Fig.~\ref{fig:prob}. As mentioned
before, these probabilities are fed into our SMPC controller at
each prediction cycle. The vertical dashed lines in both figures
stand for the moments at which the cut-in vehicles cross the
road line between two adjacent lanes. It is clear that our cut-in
detection starts around 1.25 seconds and 2 seconds ahead of
this moment for harsh and average maneuvers, respectively,
which provides a noticeable extra reaction time for the
controller.

\begin{figure*}
\centering
\begin{tabular}{ccc}
\subfigure[Spacing error in an average 5.5-sec cut-in maneuver]{\includegraphics[width=.3\textwidth]{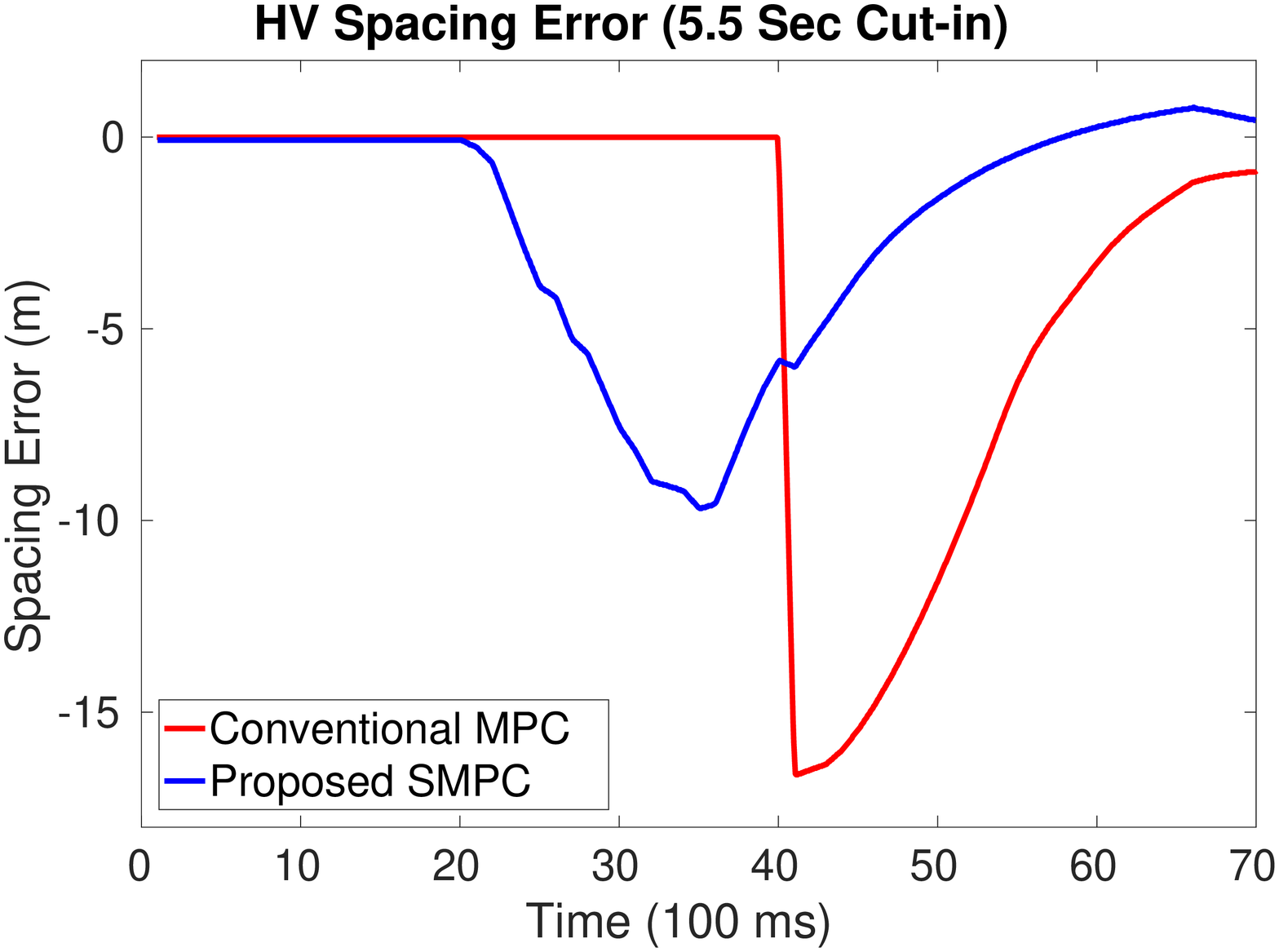}\label{fig:se5}} & 
\subfigure[Spacing error in a harsh 3-sec cut-in maneuver]{\includegraphics[width=.3\textwidth]{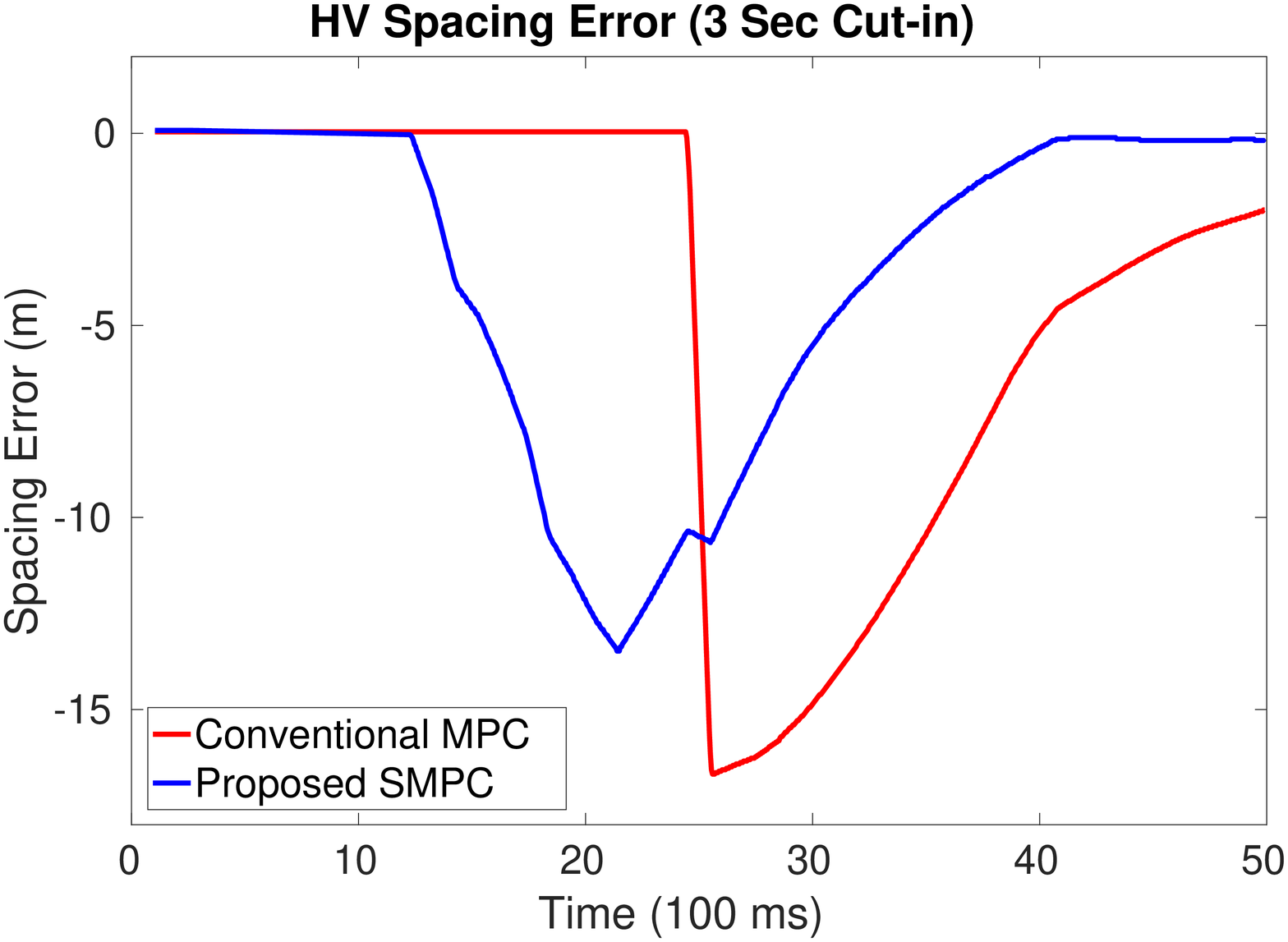}\label{fig:se3}}&
\subfigure[Velocity in an average 5.5-sec cut-in maneuver]{\includegraphics[width=.3\textwidth]{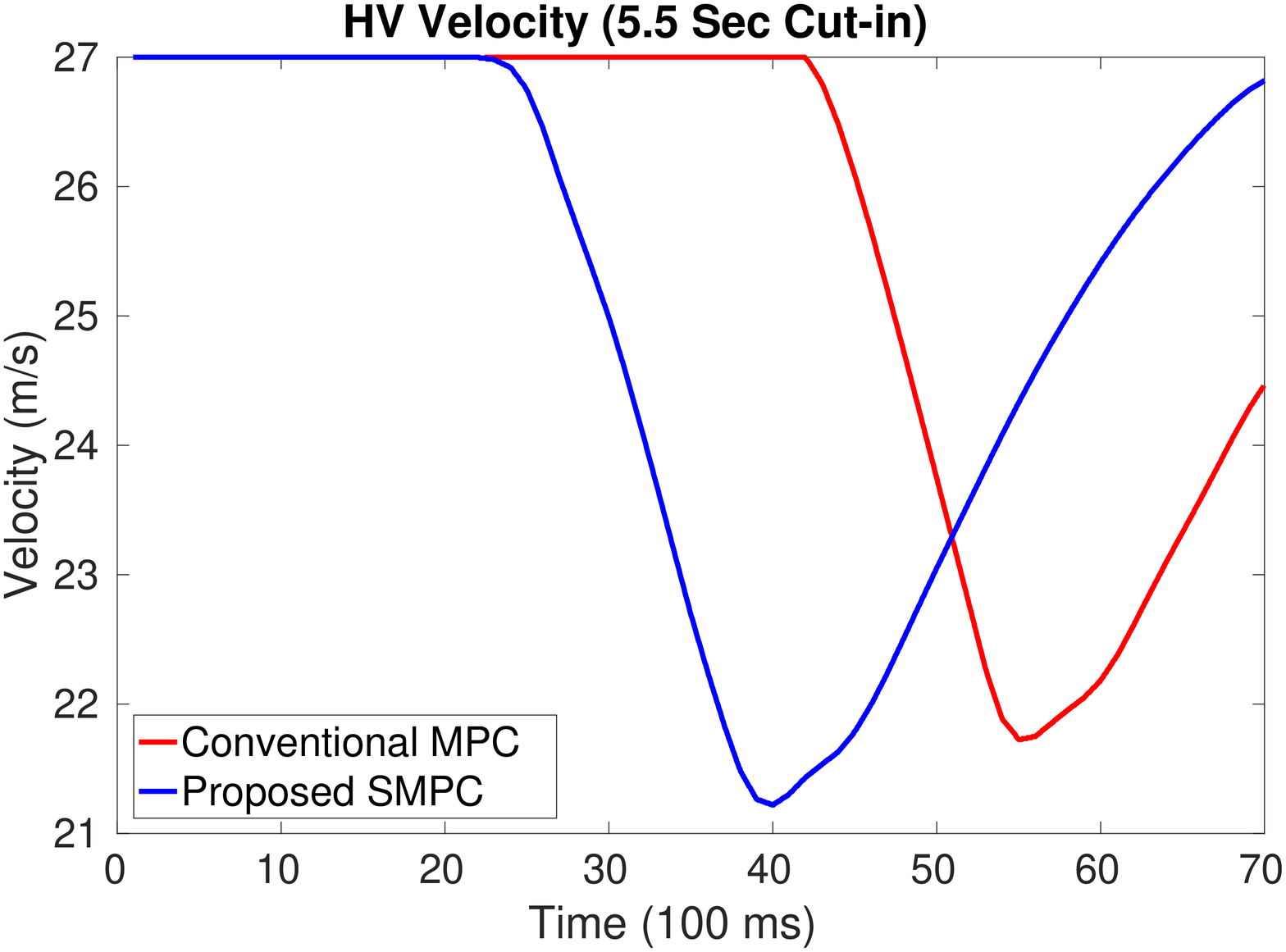}\label{fig:v5}} \\
\subfigure[Velocity in a harsh 3-sec cut-in maneuver]{\includegraphics[width=.3\textwidth]{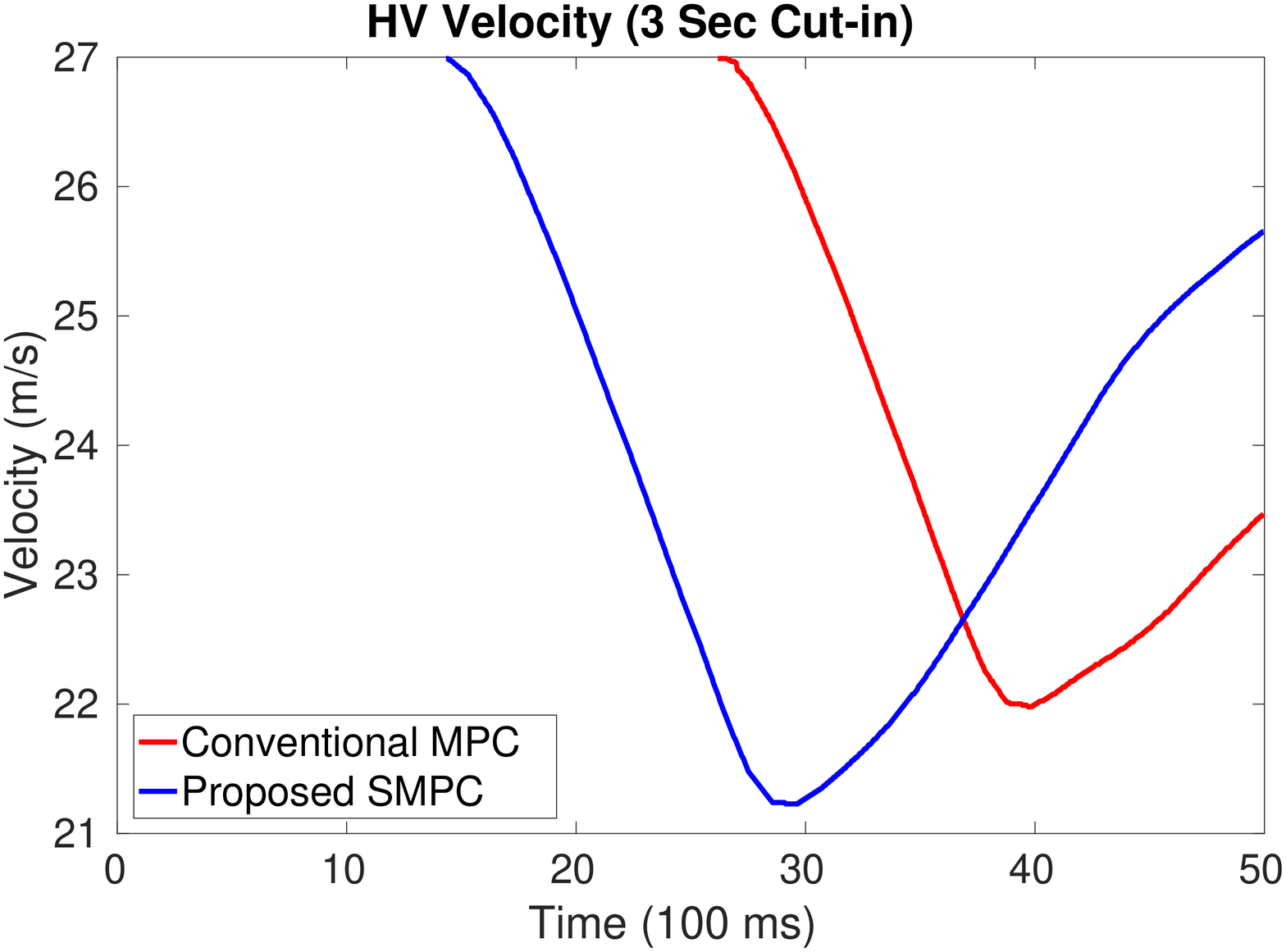}\label{fig:v3}}&
\subfigure[acceleration in an average 5.5-sec cut-in maneuver]{\includegraphics[width=.3\textwidth]{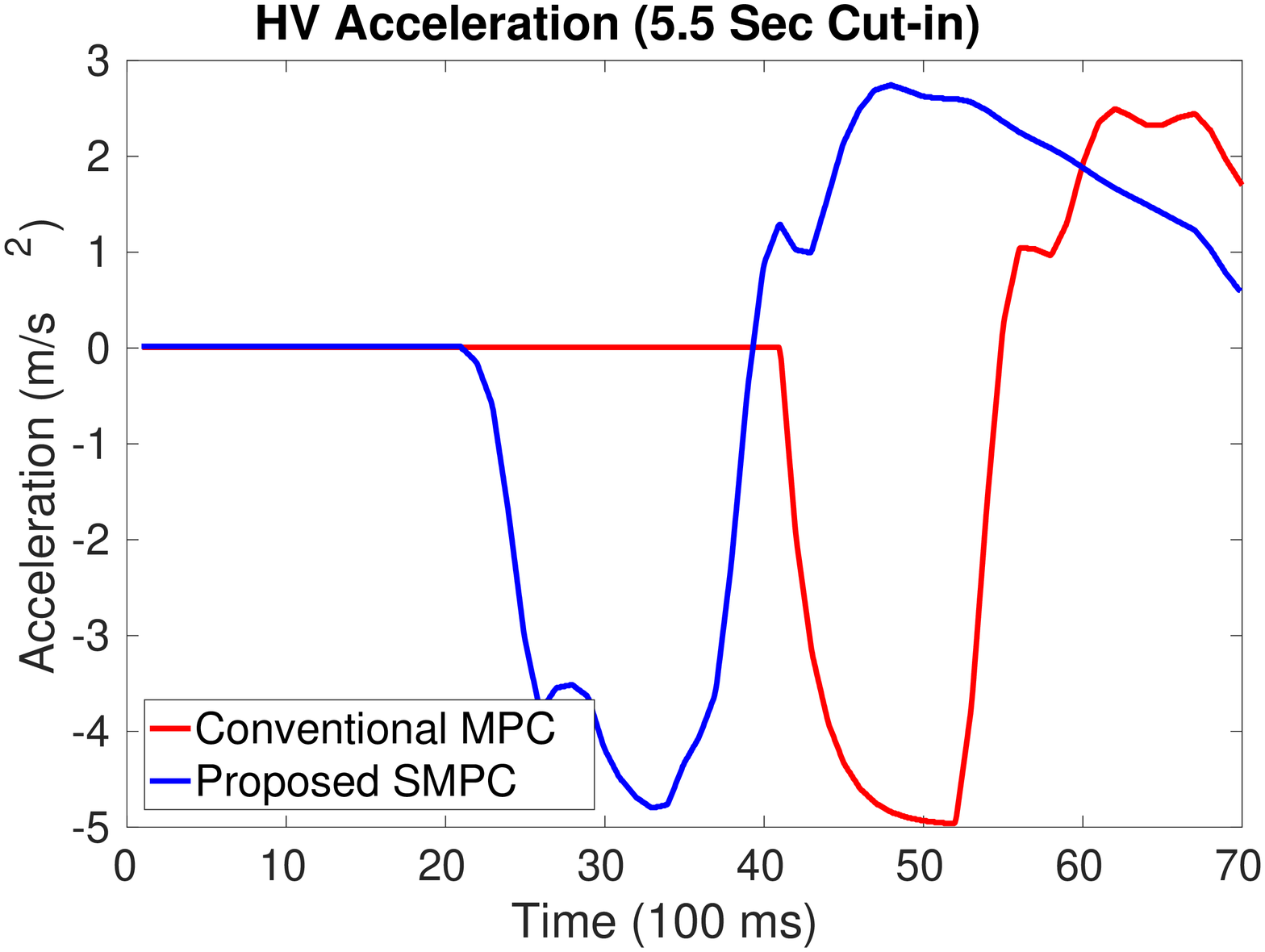}\label{fig:a5}}&
\subfigure[acceleration in a harsh 3-sec cut-in maneuver]{\includegraphics[width=.3\textwidth]{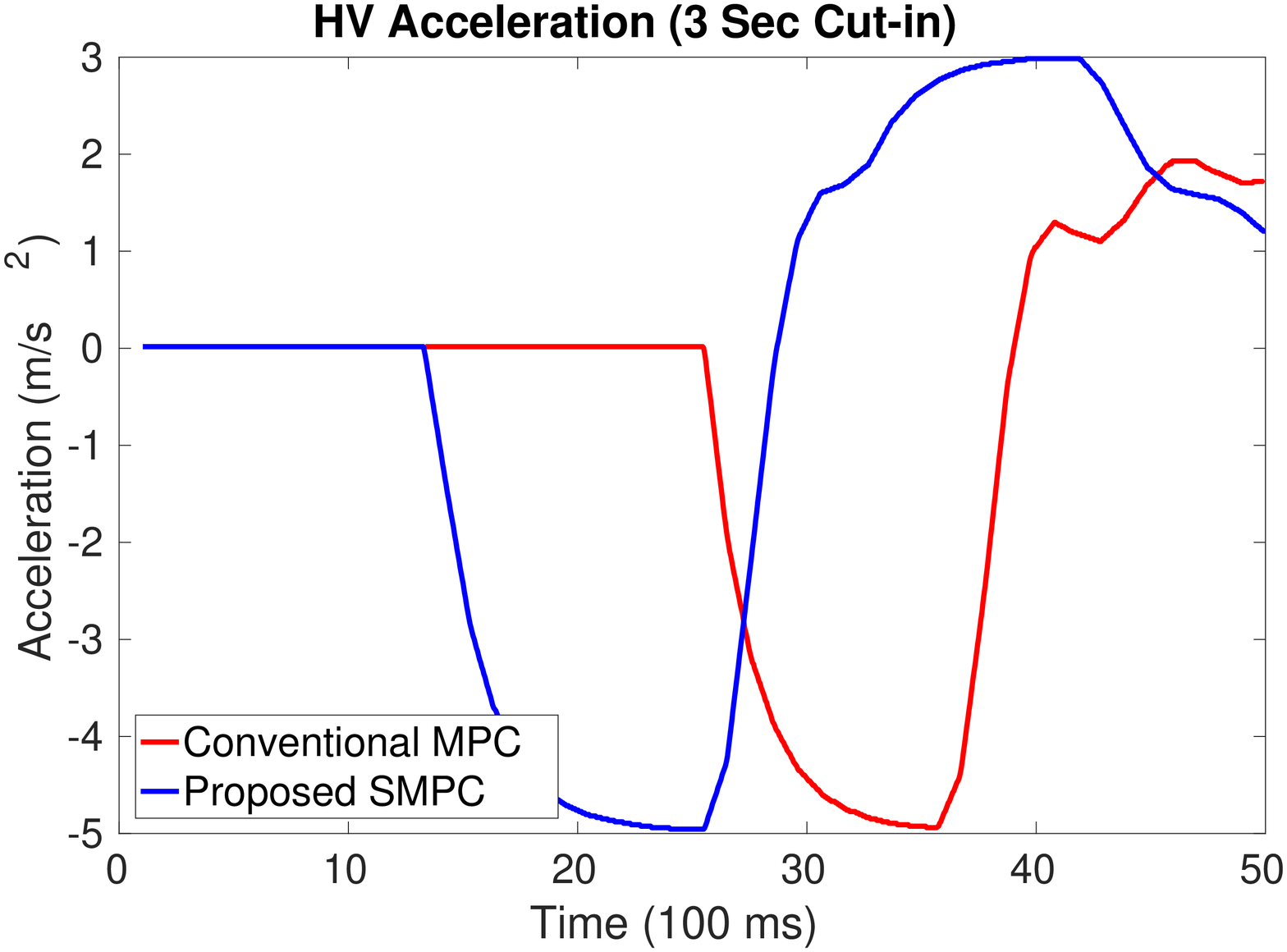}\label{fig:a3}} \\
\end{tabular}
\caption{Comparison of spacing error, velocity, and acceleration of the proposed SMPC (Blue) and the conventional MPC (Red)}
\label{fig:results}
\end{figure*}

Fig. \ref{fig:results}, illustrates the changes in spacing error,
velocity and acceleration of the host vehicle produced by two
different controllers in response to the average cut-in maneuver.
It is noteworthy that our controller starts its reaction to
compensate the situation notably sooner than the
conventional one which results in a noticeable smoother
reaction. In addition, it highly increases the reaction safety as a
consequence of its considerable lower maximum spacing error
which is evident by comparing the spacing error in Figs. \ref{fig:se5} and \ref{fig:se3}. For instance, as it is clear in Fig. \ref{fig:se5}, the worst SMPC spacing error reaches 10 meters, while its counterpart in conventional MPC system is around 17 meters. Moreover, the spacing error of the SMPC controller is around 6 meters when the suspicious vehicle entered the CACC lane while in conventional MPC it is on its maximum value of 17 meters. Finally, the SMPC cut-in detection starts around 2 seconds from the beginning of the scenario, which gives the controller about 2 seconds additional reaction time in comparison with the conventional system.

The same plots for harsh maneuver, which are depicted in
Figs. \ref{fig:se3}, \ref{fig:v3}, and \ref{fig:a3}, demonstrate the dominance of the
proposed SMPC performance in terms of sooner and safer
reaction.

\section{Conclusion} \label{sec:conc}
In this paper, a probabilistic framework for handling cut-in maneuvers into a CACC platoon is proposed and its better performance compared to the conventional controller design is demonstrated. At the first step of the designed procedure, a cut-in maneuver of an interfering vehicle is detected and its trajectory is predicted using a novel three-layer neural network-based approach. The high accuracy of this method is demonstrated by comparing its results against the state of the art Kinematic-based deterministic models. Afterwards, the output of this phase is utilized to calculate the probability of cut-in predicted trajectory overlap with the host vehicle’s bad-set area. This probability, which is referred to as cut-in probability, specifies the severity level of the dangerous situation caused by a sudden cut-in into the stable CACC platoon. Obviously, higher values of this probability need more urgent reactions from the host vehicle’s controller to prevent the possible collision with a smooth and safe reaction. This goal is achieved by giving this probability to a new stochastic MPC controller, designed based on the emerging SHS concept. The overall performance of the designed system is evaluated and its effectiveness for better regulation of the host vehicle’s reaction to dangerous cut-in situations is discussed using realistic cut-in driving scenarios from SPMD dataset.

\bibliographystyle{./IEEEtran}

\end{document}